\documentclass[preprint,aps,prd,nofootinbib,superscriptaddress,eqsecnum,tightenlines,floatfix]{revtex4-2}
\usepackage[T1]{fontenc}
\usepackage{lmodern}
\usepackage{graphicx}
\usepackage[caption=false]{subfig} 
\usepackage{amsmath,amssymb}
\usepackage{mathtools}
\usepackage{bm}

\usepackage{hyperref}
\usepackage{orcidlink}
\usepackage{zref-clever}
\usepackage{here}

\zcsetup{lang=english,cap=true,abbrev=true,nameinlink=tsingle}
\zcLanguageSetup{english}{
  namesep   = {\nobreak},
  rangesep  = {\textendash},
}
\zcRefTypeSetup{section}{
  Name-sg=Section,  Name-pl=Sections,
  Name-sg-ab=Sec.~,  Name-pl-ab=Secs.~,
}
\zcRefTypeSetup{figure}{
  Name-sg=Figure,   Name-pl=Figures,
  Name-sg-ab=Fig.~,  Name-pl-ab=Figs.~,
}
\zcRefTypeSetup{table}{
  Name-sg=Table,    Name-pl=Tables,
  Name-sg-ab=Table~, Name-pl-ab=Tables~,
}
\zcRefTypeSetup{equation}{
  Name-sg=Equation, Name-pl=Equations,
  Name-sg-ab=Eq.,   Name-pl-ab=Eqs.,
  refbounds={,(,),}
}

\DeclarePairedDelimiter{\parens}{(}{)}
\DeclarePairedDelimiter{\bracks}{[}{]}

\DeclarePairedDelimiter{\abs}{\lvert}{\rvert}

\newcommand{\del}{\partial}
\newcommand{\pd}[2]{\frac{\del #1}{\del #2}}

\newcommand{\ORCID}[1]{%
  {\,\orcidlink{#1}}%
}

\newcommand{\OK}{\text{Satisfied}}
\newcommand{\NO}{\text{Violated}}

\begin{document}

\title{Energy conditions of bouncing solutions in quadratic curvature gravity coupled with a scalar field}

\author{Yuki Hashimoto\ORCID{0000-0001-9624-2503}}
\email{s2471002@ipc.fukushima-u.ac.jp}
\affiliation{Faculty of Symbiotic Systems Science, Fukushima University, Fukushima 960-1296, Japan.}

\author{Kazuharu Bamba\ORCID{0000-0001-9720-8817}}
\email{bamba@sss.fukushima-u.ac.jp}
\affiliation{Faculty of Symbiotic Systems Science, Fukushima University, Fukushima 960-1296, Japan.}

\author{Sanjay Mandal\ORCID{0000-0003-2570-2335}}
\email{sanjaymandal960@gmail.com}
\affiliation{Faculty of Symbiotic Systems Science, Fukushima University, Fukushima 960-1296, Japan.}

\begin{abstract}
We examine the validity of classical energy conditions in nonsingular bouncing cosmological solutions arising in quadratic curvature gravity minimally coupled to a scalar field. Focusing on the null, weak, strong, and dominant energy conditions, we perform a systematic analysis under two distinct formulations of the energy-momentum tensor. In the first approach, the energy-momentum tensor is assumed to be sourced solely by the scalar field, whereas in the second, an effective energy-momentum tensor is constructed that incorporates the higher-curvature corrections characterizing deviations from general relativity. Our results reveal that, in the scalar-field description, the null, weak, and dominant energy conditions remain satisfied throughout the cosmological evolution, while the strong energy condition is necessarily violated during the bounce phase, enabling the avoidance of the initial singularity. In contrast, when the effective energy-momentum tensor is considered, all four energy conditions are violated near the bounce, highlighting the intrinsically non-Einsteinian nature of the underlying gravitational dynamics. These findings clarify the role of higher-order curvature terms in facilitating nonsingular cosmological bounces, providing important insights into the energy condition violations required in modified theories of gravity.
\end{abstract}
\maketitle
\newpage
\section{Introduction}

Inflation provides the cornerstone of modern cosmology by offering a compelling explanation for the observed large-scale homogeneity, isotropy, and spatial flatness of the Universe~\cite{Guth:1981, Sato:1981, Starobinsky:1980, Linde:1982}. In addition to resolving these classical puzzles of the standard Big Bang scenario, inflation supplies a natural mechanism for the generation of primordial quantum fluctuations, which later seed the formation of cosmic large-scale structure and the anisotropies observed in the cosmic microwave background~\cite{SatoYokoyama:2015}. Due to its remarkable explanatory power and observational success, inflation has become an indispensable component of early-universe physics.

In addition to the Planck 2018 constraints on inflation~\cite{Planck:2018jri}, the sixth data release (DR6) of the Atacama Cosmology Telescope (ACT) collaboration has recently provided high-precision CMB measurements at small angular scales~\cite{AtacamaCosmologyTelescope:2025blo}.
The ACT DR6 analysis reports a spectral index of scalar perturbations $n_s = 0.9743 \pm 0.0034$, which is slightly higher than the Planck-preferred value and can place the minimal $R^2$ model under mild tension at the $\sim 2\sigma$ level, depending on the assumptions about reheating and dataset combinations.
This development has motivated renewed investigations of extensions of minimal inflationary scenarios.
For instance, introducing a nonminimal coupling of the form $(1+\phi)R$ has been shown to reconcile the $\phi^2$ inflationary potential with the region favored by ACT DR6~\cite{Kallosh:2025rni}. More generally, the ACT-preferred tilt has led to a systematic re-examination of plateau and attractor frameworks, underscoring the importance of reheating uncertainties and the potential role of higher-curvature corrections in restoring consistency with observations~\cite{Zharov:2025, DreesXu:2025,  AddaziKetov:2025}.

The $R^2$ gravity represents the simplest extension of general relativity within the framework of $f(R)$ gravity. Although originally proposed in the context
of early-universe dynamics, $f(R)$ modifications have also been extensively studied as viable explanations for late-time cosmic acceleration  (see, e.g., Refs.~\cite{Nojiri:2006ri, padmanabhan2007, Sotiriou2008, DeFelice2010, Capozziello2011, Nojiri2010, Joyce:2014kja, Nojiri2017} for comprehensive reviews). The inflationary implications of $f(R)$ gravity have been extensively investigated~\cite{Bamba:2015uma, Odintsov:2023weg}, revealing a rich phenomenology beyond canonical scalar-field models. Nevertheless, despite its success, the inflationary paradigm does not, by itself, resolve the problem of the initial cosmological singularity implied by classical general relativity~\cite{Hawking:1970zqf}, motivating the exploration of modified gravitational frameworks capable of yielding nonsingular early-universe scenarios.

To overcome the problem of initial cosmological singularity, a wide class of nonsingular bouncing cosmologies has been proposed, in which the universe undergoes a smooth transition from a contracting phase to an expanding phase~\cite{Molina-Paris:1998xmn, Novello:2008ra, Brandenberger:2016vhg, Biswas:2005qr, Cai:2007qw, Cai:2012va, Brandenberger:2009yt}.
Various realizations of this idea have been explored, including nonsingular scenarios with nearly scale-invariant
perturbations as well as singularity-free inflationary setups~\cite{Cai:2008qw, Cai:2008qb}.
The bounce scenario is characterized kinematically in the Friedmann--Lema\^{\i}tre--Robertson--Walker (FLRW)
spacetime by the conditions that the Hubble parameter becomes zero at the bounce point
$t=t_b$, $H(t_b)=0$, and that $\dot H(t_b)>0$ holds to ensure the transition from contraction to
expansion~\cite{Molina-Paris:1998xmn, Novello:2008ra, Brandenberger:2016vhg}.
Since gravitational theory links the spacetime curvature to the energy-momentum tensor, bounce dynamics
are closely tied to the energy conditions, which are described by the attractive nature of gravity through the Raychaudhuri equations~\cite{Raychaudhuri:1953yv, Kar:2006ms, Curiel:2014zba, Visser:1997tq}.
In particular, the null energy condition (NEC) requires that matter not possess negative energy density along null
directions, and in an ideal fluid, it corresponds to the non-negativity of $\rho+P$~\cite{Curiel:2014zba, Visser:1997tq}.
Using the FLRW equations of motion, one finds that, for a flat or open universe, violating the NEC at the background
level is required in order to satisfy the bounce conditions~\cite{Molina-Paris:1998xmn, Parikh:2015bja, Battefeld:2014uga}.
By contrast, if positive spatial curvature (a closed universe) is allowed, the curvature contribution can support the
bounce, leaving room for a bounce to occur even when the matter sector satisfies the NEC~\cite{Molina-Paris:1998xmn, Battefeld:2014uga}.

Nevertheless, NEC violation often threatens the stability required for a sound gravitational theory, so the mere existence
of a bounce as a background solution is not sufficient~\cite{Battefeld:2014uga, Brandenberger:2016vhg}.
In cosmological perturbation theory, it is therefore necessary to avoid ghost instabilities associated with the sign reversal
of the kinetic term in the second-order action, as well as gradient instabilities, where short-wavelength modes grow
rapidly~\cite{Battefeld:2014uga, Kobayashi:2016xpl, Libanov:2016kfc, Carroll:2003st, Hsu:2004vr}.
Moreover, NEC-violating effective descriptions can be accompanied by additional pathologies, such as superluminality or related
instabilities, depending on the ultraviolet completion~\cite{Dubovsky:2005xd, Buniy:2005vh}.
Motivated by these issues, for NEC-violating bounces in spatially flat settings, the construction of bounce mechanisms in
Galileon theory, Horndeski theory, and their extensions have been systematically investigated, together with no-go theorems,
avoidance conditions, and stability analyses~\cite{Easson:2011zy, Qiu:2011cy, Horndeski:1974wa, Kobayashi:2016xpl, Libanov:2016kfc, Gleyzes:2014dya, Ijjas:2016vtq}.
It has also been shown that nonsingular bounces within Horndeski theory generally suffer from issues such as gradient
instabilities~\cite{Kobayashi:2016xpl, Libanov:2016kfc}, and that obtaining robust nonsingular bounces often requires extensions
beyond Horndeski~\cite{Gleyzes:2014dya, Ijjas:2016vtq}.

In modified-gravity realizations of cosmological bounces, an additional theoretical challenge arises from the potential appearance of Ostrogradsky ghosts, which typically plague non-degenerate higher-derivative theories~\cite{Woodard:2015zca}.
At the same time, it should be emphasized that $f(R)$ gravity can be rewritten as an equivalent scalar-tensor theory and thus does not
necessarily involve Ostrogradsky ghosts in this sense~\cite{Whitt:1984pd, Sotiriou2008, DeFelice2010}.
Moreover, models containing $f(R)$ face distinct theoretical constraints, such as Dolgov--Kawasaki instabilities, which must be taken into
account when discussing consistent scenarios connecting a bounce to inflation~\cite{Dolgov:2003px, Sotiriou2008, DeFelice2010}.
In addition, bouncing cosmologies have been explored in various modified-gravity frameworks, including $F(R)$ gravity, Palatini $f(R)$ gravity,
and further extensions~\cite{Bamba:2013fha, Barragan:2009sq, Odintsov:2015zza, Tripathy:2019nlw, Singh:2023gxd, Sahoo:2019qbu, Koussour:2024eig, Bhardwaj:2022xjf, Das:2025gvz}.

In this context, explicit and theoretically controlled realizations of nonsingular bounces are particularly important. A notable example is the construction of a nonsingular bouncing cosmology in a closed FLRW universe within general relativity coupled to a canonical scalar field with renormalizable self-interactions in the Jordan frame and a nonminimal curvature coupling, in which the bounce is achieved without violating the null energy condition and is instead supported by positive spatial curvature~\cite{Gungor:2020fce}.
Building on this line, a classical bouncing model in modified gravity including a quadratic-curvature $R^2$ term has been constructed, and it has been shown
that the post-bounce evolution naturally connects to an inflationary phase driven by the $R^2$ sector. A detailed stability analysis further demonstrated that the scalaron degree of freedom plays a crucial role in stabilizing the background dynamics~\cite{Daniel:2022ppp}.

These developments highlight that the formulation and interpretation of energy conditions in modified gravity is not unique, necessitating clarification of the analytical framework itself~\cite{Santos:2007bs, Bertolami:2009cd, Capozziello:2018wul, Harko:2010mv, Harko:2010zi}.
In particular, two coexisting approaches exist: (i) defining the energy condition as usual for the physical matter sector, and (ii) defining it for the ``effective
source (including the effective energy-momentum of the geometric fluid)'' obtained by reorganizing the field equations into Einstein-type form.
The interpretation of whether the NEC violation resides on the matter side or the geometry side can change depending on which approach is
adopted~\cite{Santos:2007bs, Bertolami:2009cd, Capozziello:2018wul, Harko:2010mv, Harko:2010zi}.
This point is also closely related to the geodesic convergence condition based on the Raychaudhuri equation~\cite{Raychaudhuri:1953yv, Kar:2006ms}.
Consequently, a systematic analysis of the energy condition provides an effective indicator for determining which sector bears the responsibility for satisfying the bounce
condition~\cite{Santos:2007bs, Capozziello:2018wul, Curiel:2014zba, Kar:2006ms}.

In this paper, we investigate the energy conditions of bouncing solutions in quadratic curvature gravity coupled with a scalar field.
The organization of the present work is as follows:
In \zcref{sec:energy-conditions}, we summarize how to define the effective energy-momentum tensor for the quadratic curvature gravity model with a non-minimally coupled scalar field, and how the corresponding effective energy density and pressure are related to the usual energy conditions.
In \zcref{sec:numerical-analysis}, we perform numerical analysis of the bouncing solution and evaluate the energy conditions for both the effective cosmic fluid and the scalar field solely.
Finally, conclusions are given, and the future perspectives are discussed in \zcref{sec:conclusions}.
We use units such that $c=M_{\mathrm{Pl}}=1$, where $c$ is the speed of light and $M_{\mathrm{Pl}}$ is the reduced Planck mass.
We adopt metric signature $(-,+,+,+)$.

\section{\label{sec:energy-conditions}Effective energy-momentum tensor and energy conditions}

In this section, we define an effective energy-momentum tensor $T^{(\mathrm{eff})}_{\mu\nu}$ for the quadratic curvature model with a non-minimally coupled scalar field $\varphi$ and relate the resulting effective energy density $\rho_{\mathrm{eff}}$ and pressure $P_{\mathrm{eff}}$ to the usual energy conditions.

\subsection{\label{sec:model-field-eq}Model and field equations}

In this subsection, we provide a brief review of the quadratic curvature gravity model. We also derive the equations of motion for the scalar field and the Friedmann equations, as described in Ref.~\cite{Daniel:2022ppp}.

We consider a class of $f(R,\varphi)$ theories, in which the gravitational sector is described by a function of the Ricci scalar $R$ and the scalar field $\varphi$. The action is given by
\begin{align}
  S
  = \int d^4x\,\sqrt{-g}\,
  \bigl[
    f(R,\varphi) + \mathcal{L}_{\varphi}
  \bigr],
  \label{eq:action-Jordan}
\end{align}
where $g \equiv \det(g_{\mu\nu})$ and $\mathcal{L}_{\varphi}$ denotes the Lagrangian density for $\varphi$,
\begin{align}
  \mathcal{L}_{\varphi}
  = -\frac{1}{2}(\nabla\varphi)^2 - V(\varphi),
  \label{eq:Lagrangian-phi}
\end{align}
with $V(\varphi)$ the scalar potential. Here, $\nabla_\mu$ is the covariant derivative compatible with $g_{\mu\nu}$, and we use the shorthand
$(\nabla\varphi)^2 \equiv g^{\mu\nu}\nabla_\mu\varphi\,\nabla_\nu\varphi$.

In this work, we adopt the following form
\begin{align}
  f(R,\varphi)
  = \frac{1}{2}\bigl(M_{\mathrm{Pl}}^2-\alpha\varphi^2\bigr)R
    + \frac{1}{2}A R^2,
  \label{eq:f-R-phi}
\end{align}
where $\alpha$ is a dimensionless non-minimal coupling constant and $A$ is a constant parameter controlling the magnitude of the $R^2$ curvature correction.
Here, $\psi \equiv \pd{f}{R}$ represents an extra scalar degree of freedom induced by the quadratic-curvature term, often referred to as the scalaron. From \zcref{eq:f-R-phi}, we find
\begin{align}
  \psi
  = \frac{1}{2}\bigl(M_{\mathrm{Pl}}^2-\alpha\varphi^2\bigr) + AR.
  \label{eq:R-of-psi}
\end{align}
By substituting \zcref{eq:R-of-psi} into \zcref{eq:f-R-phi}, we find that $f(R,\varphi)$ and $R$ can be cast into a bi-scalar form as follows
\begin{align}
  f
  = \psi R
    - \frac{1}{2A}\bracks*{\psi-\frac{1}{2}\bigl(M_{\mathrm{Pl}}^2-\alpha\varphi^2\bigr)}^{2},
  \quad
  R
  = \frac{1}{A}\bracks*{\psi-\frac{1}{2}\bigl(M_{\mathrm{Pl}}^2-\alpha\varphi^2\bigr)}.
  \label{eq:f-of-psi}
\end{align}
By varying the action in \zcref{eq:action-Jordan} with respect to the inverse metric $g^{\mu\nu}$, we obtain the field equation
\begin{align}
  \psi R_{\mu\nu}
  - \frac{1}{2}f\,g_{\mu\nu}
  - \parens{\nabla_\mu\nabla_\nu - g_{\mu\nu}\Box}\psi
  = T^{(\varphi)}_{\mu\nu},
  \label{eq:EOM-metric}
\end{align}
where $R_{\mu\nu}$ is the Ricci tensor and $\Box \equiv g^{\mu\nu}\nabla_\mu\nabla_\nu$ denotes the d'Alembertian operator. The energy-momentum tensor associated with $\varphi$ is defined by
\begin{align}
  T^{(\varphi)}_{\mu\nu}
  \equiv -\frac{2}{\sqrt{-g}}\,
  \frac{\delta\!\bigl(\sqrt{-g}\,\mathcal{L}_{\varphi}\bigr)}{\delta g^{\mu\nu}}
  = \nabla_\mu\varphi\,\nabla_\nu\varphi
  - g_{\mu\nu}\mathcal{L}_{\varphi}.
  \label{eq:Tmunu-phi}
\end{align}
The trace of \zcref{eq:EOM-metric} reads
\begin{equation}
  3\Box\psi + \psi R - 2f = T^{(\varphi)},
  \label{eq:trace-eq}
\end{equation}
where $T^{(\varphi)} \equiv g^{\mu\nu}T^{(\varphi)}_{\mu\nu}$. By substituting \zcref{eq:R-of-psi,eq:f-of-psi} into \zcref{eq:trace-eq}, a closed covariant evolution equation for $\psi$ is represented as
\begin{equation}
  \Box\psi
  = \frac{1}{3}\bracks*{
      \frac{M_{\mathrm{Pl}}^2-\alpha\varphi^2}{2A}
      \parens*{\psi-\frac{1}{2}\parens*{M_{\mathrm{Pl}}^2-\alpha\varphi^2}}
      + T^{(\varphi)}
    }.
  \label{eq:EOM-psi-cov}
\end{equation}
The variation of the action in \zcref{eq:action-Jordan} with respect to $\varphi$ yields the scalar-field equation of motion,
\begin{equation}
  \Box\varphi = V_{,\varphi} + \alpha\varphi R,
  \label{eq:EOM-phi-cov}
\end{equation}
where $V_{,\varphi}\equiv dV/d\varphi$.

We now consider a Friedmann--Lema\^{\i}tre--Robertson--Walker (FLRW) spacetime. In what follows, we restrict to the closed case with positive spatial curvature, $K>0$. The line element is
\begin{align}
  ds^2
  = -dt^2 + a(t)^2\bracks*{
      \frac{dr^2}{1-Kr^2} + r^2\parens*{d\theta^2+\sin^2\theta\,d\phi^2}
    },
  \label{eq:FLRW-metric}
\end{align}
where $a(t)$ is the scale factor.
The Hubble parameter is defined as $H\equiv \dot a/a$, where the dot `$\cdot$' denotes $d/dt$.

In the FLRW spacetime, the scalar field $\varphi$ can be treated as a perfect fluid with the energy density $\rho_\varphi$ and pressure $P_\varphi$ given by
\begin{equation}
  \rho_\varphi = \frac{1}{2}\dot{\varphi}^{\,2} + V(\varphi),
  \quad
  P_\varphi    = \frac{1}{2}\dot{\varphi}^{\,2} - V(\varphi).
  \label{eq:rhophi-Pphi}
\end{equation}
For this geometry, the Ricci scalar is
\begin{equation}
  R = 6\parens*{\dot H + 2H^2 + \frac{K}{a^2}}.
  \label{eq:R-FLRW}
\end{equation}
The Friedmann equations read
\begin{align}
  H^2 + \frac{K}{a^2}
  &= \frac{\psi R - f}{6\psi}
     + \frac{\rho_\varphi}{3\psi}
     - H\frac{\dot{\psi}}{\psi},
  \label{eq:Friedmann-1-mod}
\\[1mm]
  \dot{H} - \frac{K}{a^2}
  &= H\frac{\dot{\psi}}{2\psi}
   - \frac{\dot{\psi}^{\,2}}{2\psi^2}
   - \frac{\rho_\varphi + P_\varphi}{2\psi}.
  \label{eq:Friedmann-2-mod}
\end{align}
It is noted that in the FLRW spacetime, $\Box X = -\ddot X - 3H\dot X$ for any homogeneous scalar field $X(t)$ such as $\psi$ and $\varphi$. As a result, \zcref{eq:EOM-psi-cov,eq:EOM-phi-cov} can be reduced to the ordinary differential equations
\begin{align}
  \ddot\psi + 3H\dot\psi
  &= -\frac{1}{3}\bracks*{
      \frac{M_{\mathrm{Pl}}^2-\alpha\varphi^2}{2A}
      \parens*{\psi-\frac{1}{2}\parens*{M_{\mathrm{Pl}}^2-\alpha\varphi^2}}
      + T^{(\varphi)}
    },
\label{eq:EOM-psi-FLRW}
\\[1mm]
  \ddot\varphi + 3H\dot\varphi
  &= -V_{,\varphi} - \alpha\varphi R,
\label{eq:EOM-phi-FLRW}
\end{align}
with $R$ given by \zcref{eq:R-FLRW}.

\subsection{\label{subsec:teff}Definition of the effective energy-momentum tensor}
In this subsection, by introducing an effective\footnote{Here, the term ``effective'' refers to quantities obtained by formally recasting the gravitational field equations in the quadratic curvature theory into the following
Einstein-like form $G_{\mu\nu} = M_{\mathrm{Pl}}^{-2} T^{(\mathrm{eff})}_{\mu\nu}$, as in \zcref{eq:Einstein-eff}.
In this representation, $T^{(\mathrm{eff})}_{\mu\nu}$ does not describe
the physical energy-momentum tensor of the matter field alone, but includes
geometric contributions arising from the higher-curvature terms and the
non-minimal coupling to the scalar field. These contributions can be
interpreted as the effective cosmic fluid that reproduces the same spacetime dynamics
as the original gravitational system described by the action in \zcref{eq:action-Jordan}, while for the matter energy-momentum tensor $T^{(\varphi)}_{\mu\nu}$, the usual energy conditions can be satisfied.} energy-momentum tensor $T^{(\mathrm{eff})}_{\mu\nu}$, we recast the gravitational field equations into the following Einstein-like form
\begin{align}
  G_{\mu\nu} = \frac{1}{M_{\mathrm{Pl}}^2}\,T^{(\mathrm{eff})}_{\mu\nu}.
  \label{eq:Einstein-eff}
\end{align}
Here, $G_{\mu\nu}$ is the Einstein tensor. Since $\nabla^\mu G_{\mu\nu}=0$ identically (the Bianchi identity), ${T^{(\mathrm{eff})}_{\mu\nu}}$ is covariantly conserved as $\nabla^\mu {T^{(\mathrm{eff})}_{\mu\nu}}=0$, by construction.
Equivalently, we define
\begin{align}
  T^{(\mathrm{eff})}_{\mu\nu} \equiv M_{\mathrm{Pl}}^2\,G_{\mu\nu}.
  \label{eq:Teff-definition}
\end{align}
For the closed, homogeneous, and isotropic background introduced above, the non-zero components of the Einstein tensor are given by
\begin{align}
  G^0{}_0 &= -3\parens*{H^2+\frac{K}{a^2}},
  \\
  G^i{}_j &= -\parens*{2\dot H + 3H^2 + \frac{K}{a^2}}\delta^i{}_j.
\end{align}
Motivated by the perfect-fluid form ${T^\mu}_\nu=\mathrm{diag}(-\rho,\,P,\,P,\,P)$, we define the effective energy density $\rho_{\mathrm{eff}}$ and pressure $P_{\mathrm{eff}}$ through
\begin{align}
  G^0{}_0 = -\frac{\rho_{\mathrm{eff}}}{M_{\mathrm{Pl}}^2},
  \quad
  G^i{}_j = \frac{P_{\mathrm{eff}}}{M_{\mathrm{Pl}}^2}\delta^i{}_j,
\end{align}
which immediately yields
\begin{align}
  \rho_{\mathrm{eff}}
  &= 3M_{\mathrm{Pl}}^2\parens*{H^2+\frac{K}{a^2}},
  \label{eq:rho-eff-def}
\\[1mm]
  P_{\mathrm{eff}}
  &= -M_{\mathrm{Pl}}^2\parens*{2\dot H + 3H^2+\frac{K}{a^2}}.
  \label{eq:P-eff-def}
\end{align}
For later convenience, we also record the following combinations
\begin{align}
  \rho_{\mathrm{eff}} + P_{\mathrm{eff}}
  &= -2M_{\mathrm{Pl}}^2\parens*{\dot H - \frac{K}{a^2}},
  \label{eq:rho+P-eff}
\\
  \rho_{\mathrm{eff}} + 3P_{\mathrm{eff}}
  &= -6M_{\mathrm{Pl}}^2\parens*{\dot H + H^2}.
  \label{eq:rho+3P-eff}
\end{align}

\subsection{\label{subsec:consistency}Consistency with the modified Friedmann equations}
In this subsection, we show that the similar representations to general relativity, \zcref{eq:rho-eff-def,eq:P-eff-def} are consistent with the underlying Friedmann equations \eqref{eq:Friedmann-1-mod} and \eqref{eq:Friedmann-2-mod}.
By using \zcref{eq:Friedmann-1-mod}, \zcref{eq:rho-eff-def} can be rewritten as
\begin{align}
  \rho_{\mathrm{eff}}
  &= 3M_{\mathrm{Pl}}^2\parens*{H^2+\frac{K}{a^2}}
  \nonumber\\
  &= M_{\mathrm{Pl}}^2\biggl[
      \frac{\psi R - f}{2\psi}
      + \frac{\rho_\varphi}{\psi}
      - 3H\frac{\dot\psi}{\psi}
     \biggr].
\label{eq:rho-eff-from-mod}
\end{align}
Similarly, combining \zcref{eq:rho+P-eff} with \zcref{eq:Friedmann-2-mod} yields
\begin{align}
  \rho_{\mathrm{eff}} + P_{\mathrm{eff}}
  &= -2M_{\mathrm{Pl}}^2\parens*{\dot H - \frac{K}{a^2}}
  \nonumber\\
  &= -M_{\mathrm{Pl}}^2\biggl[
      H\frac{\dot\psi}{\psi}
      - \frac{\dot\psi^{\,2}}{\psi^2}
      - \frac{\rho_\varphi + P_\varphi}{\psi}
     \biggr].
\label{eq:rho+P-eff-from-mod}
\end{align}
These relations show that the effective energy density described by \zcref{eq:rho-eff-def} and effective pressure described by \zcref{eq:P-eff-def} are consistent with the underlying gravitational field equations.

\subsection{\label{subsec:energy-conditions}Energy conditions}
In this subsection, we relate the effective energy density $\rho_{\mathrm{eff}}$ and pressure $P_{\mathrm{eff}}$ defined in \zcref{subsec:teff} to the standard energy conditions.
For a perfect fluid with the energy density $\rho$ and pressure $P$, the standard energy conditions are summarized as follows.
\begin{itemize}
  \item \textbf{Null energy condition (NEC):} $\rho + P \ge 0$.
  \item \textbf{Strong energy condition (SEC):} $\rho + 3P \ge 0$.
  \item \textbf{Weak energy condition (WEC):} $\rho \ge 0$ and $\rho + P \ge 0$.
  \item \textbf{Dominant energy condition (DEC):} $\rho \ge 0$ and $\rho \ge \abs{P}$.
\end{itemize}
By utilizing \zcref{eq:rhophi-Pphi}, the energy conditions for the scalar field alone are expressed as follows
\begin{align}
  \text{NEC (scalar field)} \quad
  &\Longleftrightarrow\quad
  \rho_\varphi + P_\varphi = \dot\varphi^{\,2} \ge 0,
  \label{eq:nec-scalar}
\\[1mm]
\text{SEC (scalar field)} \quad
  &\Longleftrightarrow\quad
  \rho_\varphi + 3P_\varphi = 2\dot\varphi^{\,2} - 2V\parens{\varphi} \ge 0,
  \label{eq:sec-scalar}
\\[1mm]
  \text{WEC (scalar field)} \quad
  &\Longleftrightarrow\quad
  \rho_\varphi \ge 0,
  \quad
  \rho_\varphi + P_\varphi \ge 0,
  \label{eq:wec-scalar}
\\[1mm]
  \text{DEC (scalar field)} \quad
  &\Longleftrightarrow\quad
  \rho_\varphi \ge 0,
  \quad
  \rho_\varphi \ge \abs{P_\varphi}.
  \label{eq:dec-scalar}
\end{align}
Similarly, by using \zcref{eq:rho-eff-def,eq:rho+3P-eff}, the energy-conditions for the effective cosmic fluid become
\begin{align}
  \text{NEC (effective)} \quad
  &\Longleftrightarrow\quad
  \rho_{\mathrm{eff}} + P_{\mathrm{eff}}
  = -2M_{\mathrm{Pl}}^2\parens*{\dot H-\frac{K}{a^2}} \ge 0,
  \label{eq:nec-eff}
\\[1mm]
\text{SEC (effective)} \quad
  &\Longleftrightarrow\quad
  \rho_{\mathrm{eff}} + 3P_{\mathrm{eff}}
  = -6M_{\mathrm{Pl}}^2\parens*{\dot H + H^2} \ge 0,
  \label{eq:sec-eff}
\\[1mm]
  \text{WEC (effective)} \quad
  &\Longleftrightarrow\quad
  \rho_{\mathrm{eff}} \ge 0,
  \quad
  \rho_{\mathrm{eff}} + P_{\mathrm{eff}} \ge 0,
  \label{eq:wec-eff}
\\[1mm]
  \text{DEC (effective)} \quad
  &\Longleftrightarrow\quad
  \rho_{\mathrm{eff}} \ge 0,
  \quad
  \rho_{\mathrm{eff}} \ge \abs{P_{\mathrm{eff}}}.
  \label{eq:dec-eff}
\end{align}

\section{\label{sec:numerical-analysis}Evolution of the bounce dynamics and energy conditions}
In this section, we present numerical solutions of the background dynamics and evaluate the energy conditions defined in \zcref{sec:energy-conditions}.

We numerically solve the coupled background dynamics in the FLRW spacetime, consisting of the Friedmann equations \eqref{eq:Friedmann-1-mod} and \eqref{eq:Friedmann-2-mod} and the equations of motion for the scalar field \eqref{eq:EOM-psi-FLRW} and \eqref{eq:EOM-phi-FLRW}, with $R$ given by \zcref{eq:R-FLRW}.
The time derivative of the scale factor evolves as
\begin{equation}
  \dot a = aH.
  \label{eq:adot}
\end{equation}

\subsection{Model parameters and initial conditions}

We consider the potential of the scalar field $\varphi$ given by~\cite{Gungor:2020fce,Daniel:2022ppp}
\begin{equation}
  V\parens{\varphi}
  = \frac{1}{2}m^2\varphi^2 + \frac{1}{3}\beta\varphi^3 + \frac{1}{4}\lambda\varphi^4.
\label{eq:potential}
\end{equation}
This potential, combined with the non-minimal coupling to gravity, can support a nonsingular bounce in a closed FLRW universe without violating the null energy condition for the scalar field alone~\cite{Gungor:2020fce,Daniel:2022ppp}.

We adopt the same parameter values as in Ref.~\cite{Daniel:2022ppp}:
$M_{\mathrm{Pl}}=1$, $m=10^{-6}M_{\mathrm{Pl}}$, $\lambda=10^{-12}$, $\beta_{\mathrm{fac}}=4.49$ with $\beta=-\sqrt{\beta_{\mathrm{fac}}\lambda}\,m$, $\alpha=10^{-3}$, $A=10^{12}M_{\mathrm{Pl}}^{-2}$, $K=m^2$, and $a_{\mathrm{i}}=10^2$.
Here, $m$ is the mass parameter in the scalar potential, $\lambda$ is the quartic coupling, $\beta$ is the cubic coupling, $\alpha$ is the non-minimal coupling constant, $A$ controls the strength of the $R^2$ term, $K>0$ sets positive spatial curvature, and $a_{\mathrm{i}}$ is the initial value of the scale factor.

For simplification, we also define the non-minimal coupling function as
\begin{equation}
  F\parens{\varphi} \equiv M_{\mathrm{Pl}}^2 - \alpha\varphi^2.
\label{eq:def-F}
\end{equation}

At the initial time $t=t_{\mathrm{i}}$, we set $\varphi(t_{\mathrm{i}})=\varphi_\mathrm{i} = \frac{-\beta + \sqrt{\beta^2 -4\lambda m^2}}{2\lambda}$ and impose $\psi(t_{\mathrm{i}})=\psi_{\mathrm{i}}$ with
$\psi_{\mathrm{i}}=\frac{8A\,V\parens{\varphi_{\mathrm{i}}}}{F\parens{\varphi_{\mathrm{i}}}}+\frac{1}{2}F\parens{\varphi_{\mathrm{i}}}$,
$H(t_{\mathrm{i}})=H_{\mathrm{i}}$ with $H_{\mathrm{i}}^2=\frac{V\parens{\varphi_{\mathrm{i}}}}{3F\parens{\varphi_{\mathrm{i}}}}$ and $H_{\mathrm{i}}<0$ (selecting an initially contracting branch), and $\dot\varphi(t_{\mathrm{i}})=0$, $\dot\psi(t_{\mathrm{i}})=0$, and $a(t_{\mathrm{i}})=a_{\mathrm{i}}$.

We introduce the dimensionless time variable
\begin{equation}
  \tau \equiv m\parens{t-t_{\mathrm{b}}},
\label{eq:tau-def}
\end{equation}
where $t_{\mathrm{b}}$ denotes the bounce time defined by $H(t_{\mathrm{b}})=0$.
Thus, when $\tau=0$, the bounce occurs.

\subsection{Bounce dynamics}

\begin{figure}[tbp]
  \includegraphics[width=\linewidth]{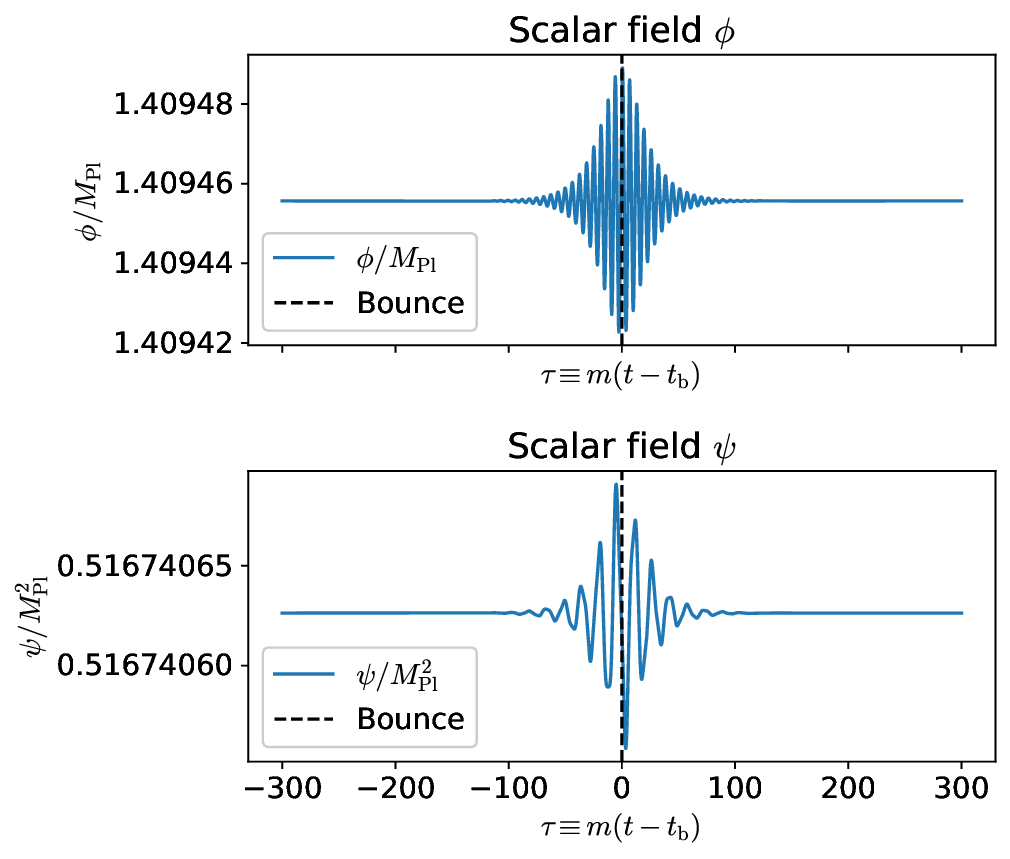}
  \caption{\label{fig:scalar_fields}Evolution of the scalar fields in the FLRW spacetime.
  The horizontal axis is the dimensionless time $\tau\equiv m(t-t_{\mathrm{b}})$, and the vertical dashed line marks the bounce at $\tau=0$.
  Top panel: $\varphi/M_{\mathrm{Pl}}$ as a function of $\tau$ (solid curve).
  Bottom panel: $\psi/M_{\mathrm{Pl}}^{2}$ as a function of $\tau$ (solid curve).
  We use the parameter set
  $M_{\mathrm{Pl}}=1$, $m=10^{-6}M_{\mathrm{Pl}}$, $\lambda=10^{-12}$, $\beta_{\mathrm{fac}}=4.49$, $\beta=-\sqrt{\beta_{\mathrm{fac}}\lambda}\,m$, $\alpha=10^{-3}$, $A=10^{12}M_{\mathrm{Pl}}^{-2}$, $K=m^2$, $a_{\mathrm{i}}=10^2$
  and the initial conditions
  $\varphi_\mathrm{i} = \frac{-\beta + \sqrt{\beta^2 -4\lambda m^2}}{2\lambda}$, $\psi_{\mathrm{i}}=\frac{8A\,V(\varphi_{\mathrm{i}})}{F(\varphi_{\mathrm{i}})}+\frac{1}{2}F(\varphi_{\mathrm{i}})$,
  $H_{\mathrm{i}}^2=\frac{V(\varphi_{\mathrm{i}})}{3F(\varphi_{\mathrm{i}})}$ with $H_{\mathrm{i}}<0$,
  $\dot\varphi_{\mathrm{i}}=0$, $\dot\psi_{\mathrm{i}}=0$, and $a(t_{\mathrm{i}})=a_{\mathrm{i}}$,
  where $V(\varphi)$ is given by \zcref{eq:potential} and $F(\varphi)=M_{\mathrm{Pl}}^2-\alpha\varphi^2$.}
\end{figure}

In \zcref{fig:scalar_fields}, we show the time evolution of the scalar field $\varphi$ and the scalaron $\psi$ around the bounce time, $\tau=0$.
Both fields remain finite and nearly constant away from the bounce, while exhibiting localized oscillatory features near $\tau\simeq 0$.
In particular, $\varphi$ undergoes a small-amplitude transient oscillation superposed on an approximately constant background value, and $\psi$ shows a corresponding oscillatory response.
These correlated oscillations indicate an energy exchange between the matter sector and the curvature-induced scalar degree of freedom, triggered in the high-curvature regime around the bounce.
The rapid decay of these oscillations away from $\tau=0$ is consistent with the damping expected from the Hubble friction terms in \zcref{eq:EOM-psi-FLRW,eq:EOM-phi-FLRW}.

\begin{figure}[tbp]
  \includegraphics[width=\linewidth]{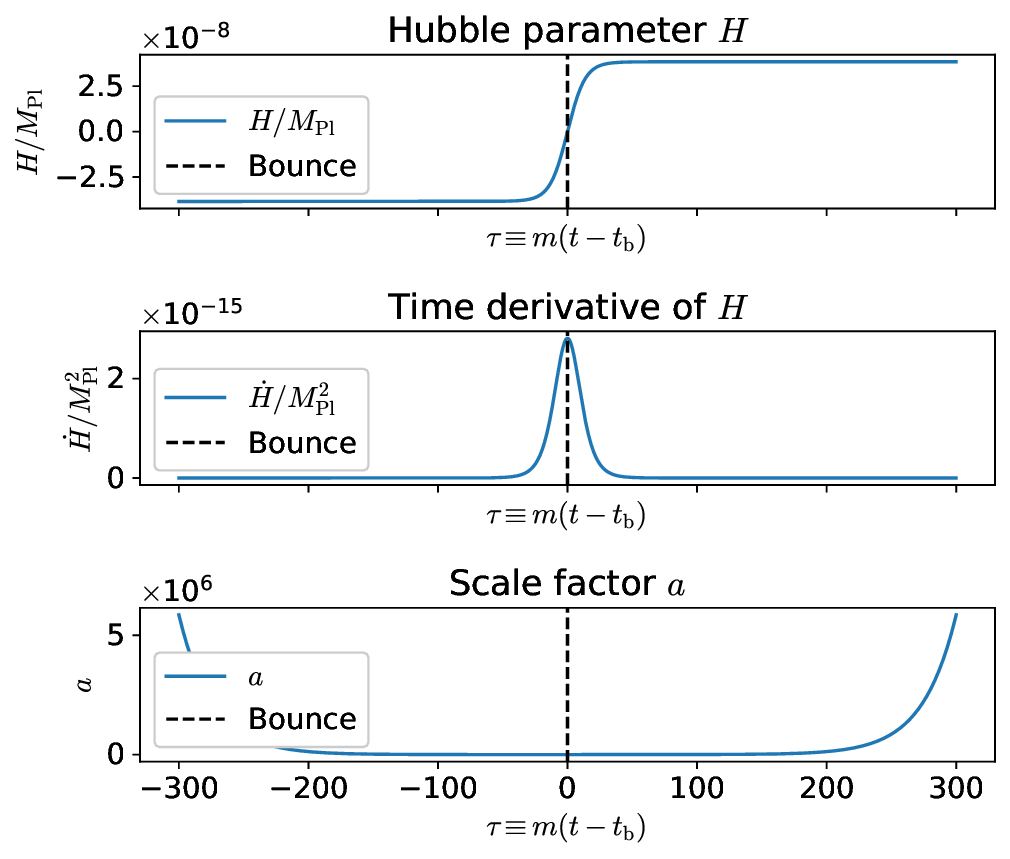}
  \caption{\label{fig:hubble_parameters}Evolution of the background variables using the same parameters and initial conditions as in \zcref{fig:scalar_fields}.
  The horizontal axis is $\tau\equiv m(t-t_{\mathrm{b}})$, and the vertical dashed line indicates the bounce ($\tau=0$).
  Top panel: $H/M_{\mathrm{Pl}}$ as a function of $\tau$ (solid curve).
  Middle panel: $\dot H/M_{\mathrm{Pl}}^{2}$ as a function of $\tau$ (solid curve), where the dot `$\cdot$' denotes $d/dt$.
  Bottom panel: $a$ as a function of $\tau$ (solid curve).}
\end{figure}

The background evolution of the scalar fields, the Hubble parameter, the time derivative of the Hubble parameter, and the scale factor is summarized in \zcref{fig:hubble_parameters}.
The Hubble parameter evolves smoothly from negative values (contraction) to positive values (expansion) and crosses zero at $\tau=0$, in accordance with the definition of the bounce time $t_{\mathrm{b}}$.
Moreover, $\dot H$ becomes positive around $\tau=0$ and exhibits a pronounced peak at the bounce, which ensures that the scale factor attains a local minimum and the transition from contraction to expansion is nonsingular.
Correspondingly, the scale factor decreases during the contracting phase, reaches its minimum at the bounce, and subsequently increases in the expanding phase.
After the bounce, $a(t)$ grows rapidly while $H$ remains nearly constant and positive, which is consistent with an accelerated expansion phase.

\subsection{Equation of state}

\begin{figure}[tbp]
  \includegraphics[width=\linewidth]{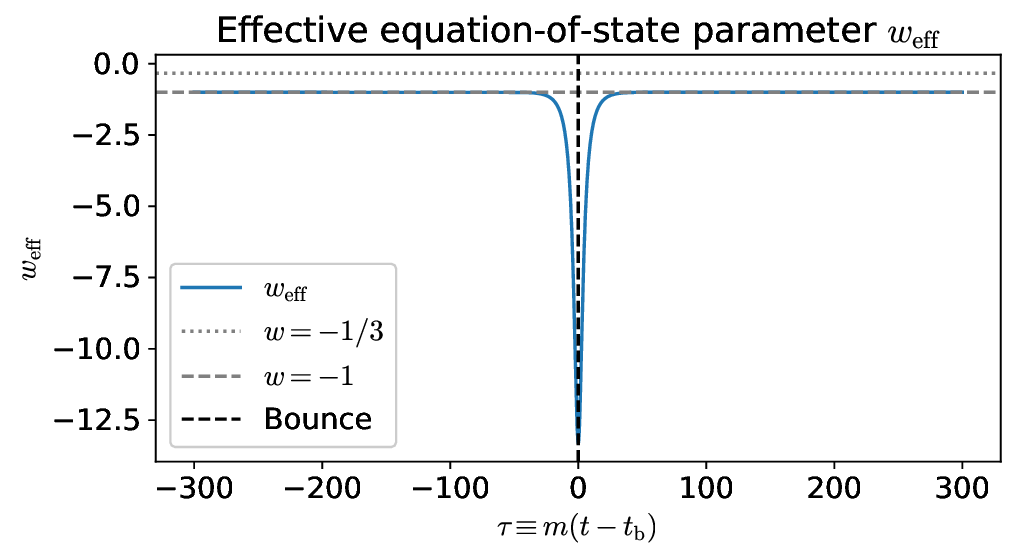}
  \caption{\label{fig:w_t}Evolution of $w_{\mathrm{eff}}(\tau)$ defined in \zcref{eq:w-eff-def}, using the same parameters and initial conditions as in \zcref{fig:scalar_fields}.
  The horizontal axis is $\tau\equiv m(t-t_{\mathrm{b}})$, and the vertical dashed line indicates the bounce ($\tau=0$).
  The solid curve shows $w_{\mathrm{eff}}$.
  The horizontal reference lines show $w=-1$ (dashed) and $w=-1/3$ (dotted).}
\end{figure}

We define the effective equation-of-state parameter by
\begin{equation}
  w_{\mathrm{eff}} \equiv \frac{P_{\mathrm{eff}}}{\rho_{\mathrm{eff}}},
  \label{eq:w-eff-def}
\end{equation}
where $\rho_{\mathrm{eff}}$ and $P_{\mathrm{eff}}$ are given by \zcref{eq:rho-eff-def,eq:P-eff-def}.
The resulting evolution is presented in \zcref{fig:w_t}.
Away from the bounce, $w_{\mathrm{eff}}$ stays close to $-1$, consistent with an approximately de Sitter-like background.
In contrast, the bounce phase features a sharp negative excursion of $w_{\mathrm{eff}}$.
This transient ``phantom-like'' behavior reflects that, once the modified-gravity dynamics are recast into the Einstein-like form \zcref{eq:Einstein-eff}, the corresponding effective cosmic fluid violates the null-energy bound near the bounce (see also \zcref{fig:effective-energy-conditions}).
After the bounce, $w_{\mathrm{eff}}$ rapidly returns to its quasi-de Sitter value.

\subsection{Energy conditions}

In this subsection, we evaluate the energy conditions for (i) the scalar field sector solely defined by \zcref[range]{eq:nec-scalar,eq:dec-scalar} and (ii) the effective cosmic fluid defined by \zcref[range]{eq:nec-eff,eq:dec-eff}.

\begin{figure}[tbp]
  \centering
  \subfloat[NEC: $\rho_\varphi+P_\varphi$ \label{fig:scalar-field-energy-conditions-nec}]{
    \includegraphics[width=0.48\linewidth]{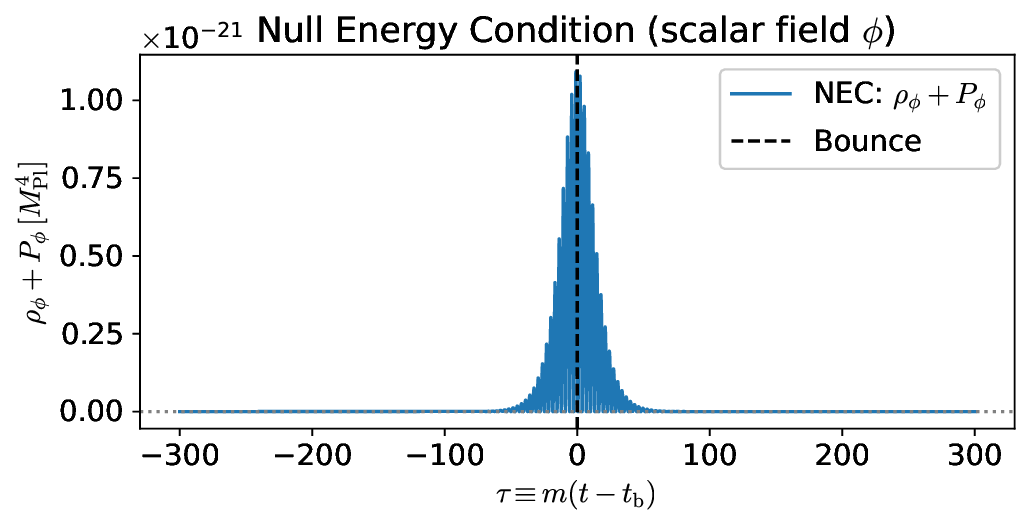}
  }\hfill
  \subfloat[SEC: $\rho_\varphi+3P_\varphi$ \label{fig:scalar-field-energy-conditions-sec}]{
    \includegraphics[width=0.48\linewidth]{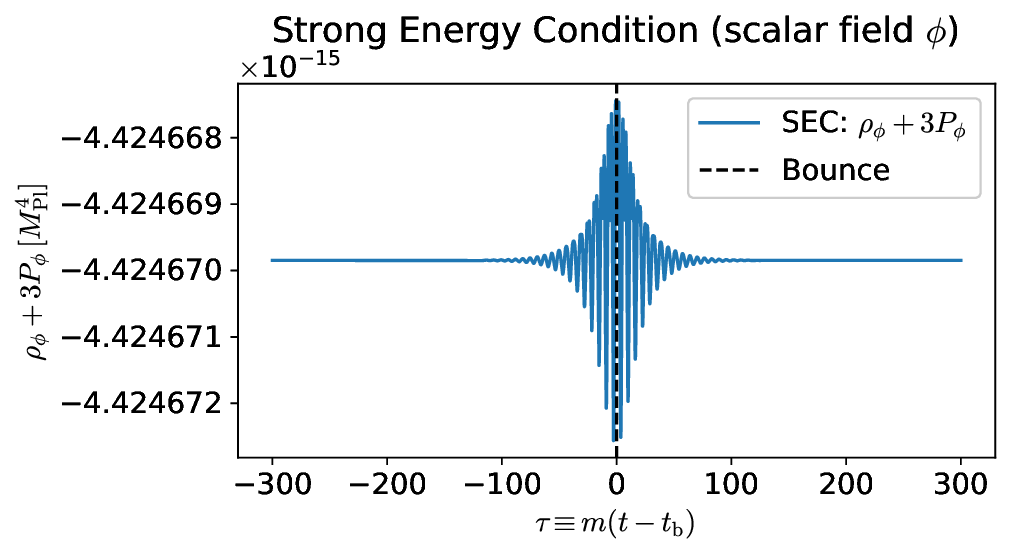}
  }\\[0.8ex]
  \subfloat[WEC: $\rho_\varphi$ and $\rho_\varphi+P_\varphi$ \label{fig:scalar-field-energy-conditions-wec}]{
    \includegraphics[width=0.48\linewidth]{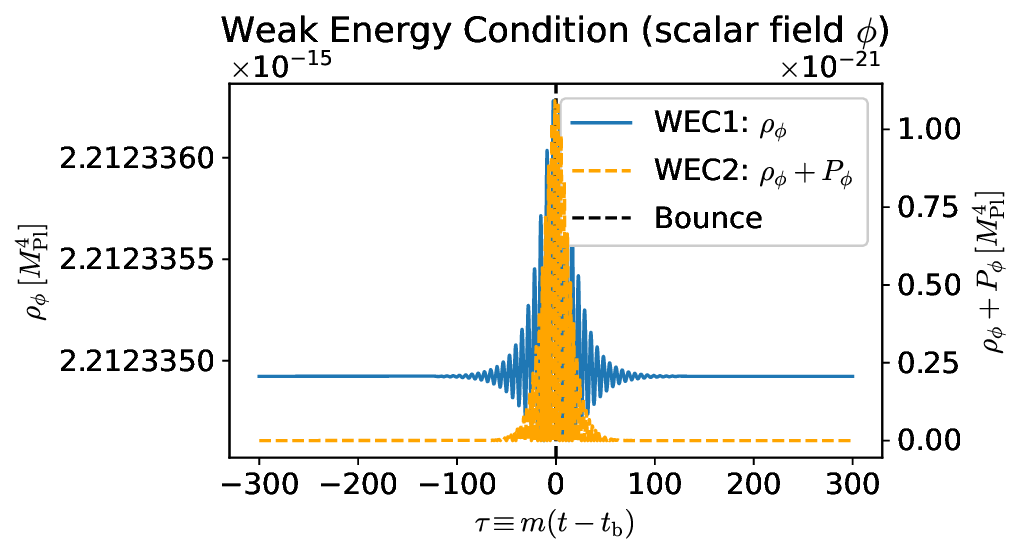}
  }\hfill
  \subfloat[DEC: $\rho_\varphi - \abs{P_\varphi}$ \label{fig:scalar-field-energy-conditions-dec}]{
    \includegraphics[width=0.48\linewidth]{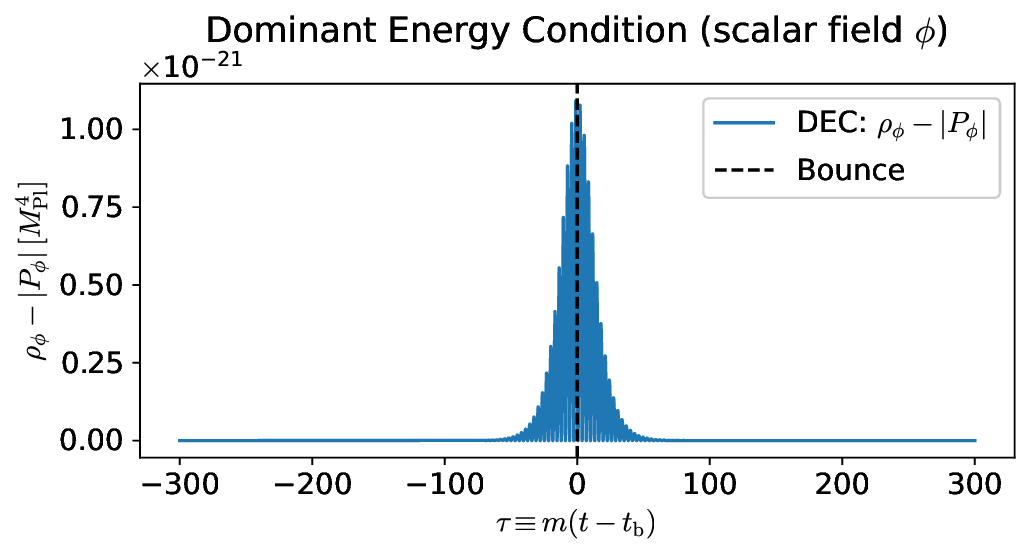}
  }
  \caption{\label{fig:scalar-field-energy-conditions}Evolution of the scalar-field energy-condition indicators, using the same parameters and initial conditions as in \zcref{fig:scalar_fields}.
  The horizontal axis in all panels is $\tau\equiv m(t-t_{\mathrm{b}})$, and the vertical dashed line marks the bounce ($\tau=0$).
  Panels (a)--(d) correspond to NEC, SEC, WEC, and DEC diagnostics, respectively.
  All quantities are normalized as indicated on the plot axes.}
\end{figure}

The corresponding energy conditions for the scalar field are shown in \zcref{fig:scalar-field-energy-conditions}.
As expected from \zcref{eq:rhophi-Pphi}, the null-energy combination $\rho_\varphi+P_\varphi=\dot\varphi^{\,2}$ is nonnegative, and both $\rho_\varphi$ and $\rho_\varphi+P_\varphi$ remain nonnegative, so that NEC and WEC hold for the scalar sector.
The dominant-energy indicator is also nonnegative, implying DEC satisfaction.
On the other hand, $\rho_\varphi+3P_\varphi=2\dot\varphi^{\,2}-2V(\varphi)$ becomes negative around the bounce, demonstrating SEC violation.
This behavior is consistent with the well-known requirement that a nonsingular bounce typically necessitates SEC violation at the background level.
The overall status at $\tau=0$ is summarized in \zcref{tab:energy_conditions_bounce}.

\begin{figure}[tbp]
  \centering
  \subfloat[NEC: $\rho_{\mathrm{eff}}+P_{\mathrm{eff}}$ \label{fig:effective-energy-conditions-nec}]{
    \includegraphics[width=0.48\linewidth]{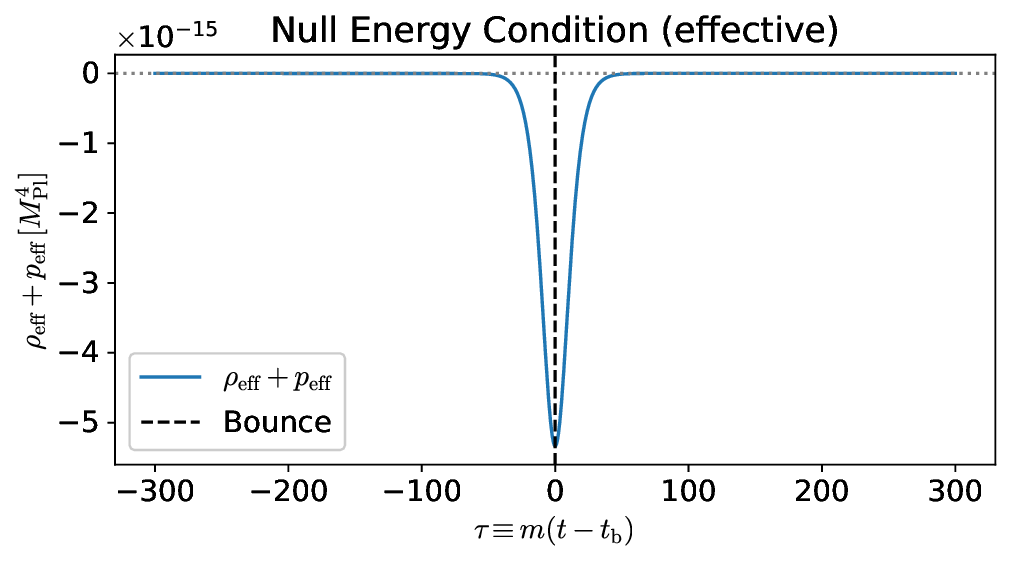}
  }\hfill
  \subfloat[SEC: $\rho_{\mathrm{eff}}+3P_{\mathrm{eff}}$ \label{fig:effective-energy-conditions-sec}]{
    \includegraphics[width=0.48\linewidth]{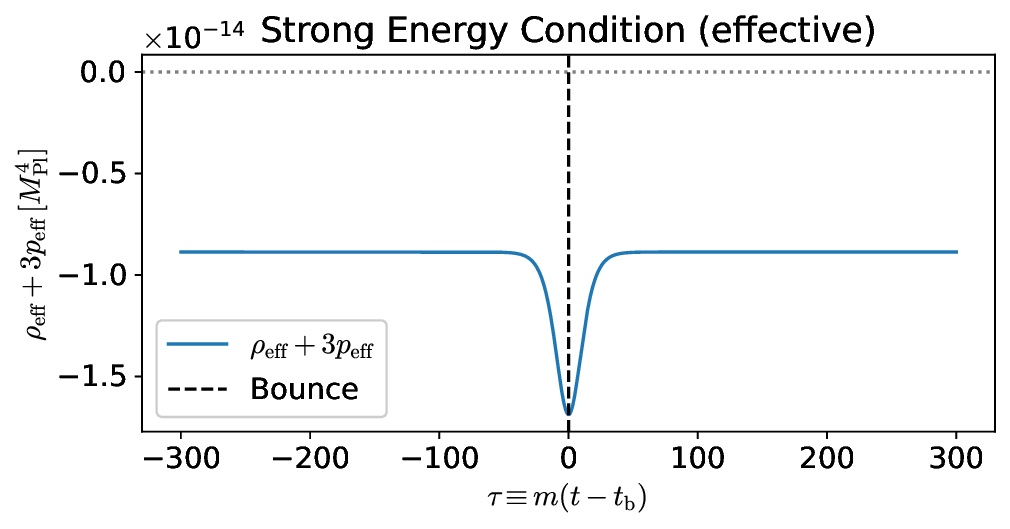}
  }\\[0.8ex]
  \subfloat[WEC: $\rho_{\mathrm{eff}}$ and $\rho_{\mathrm{eff}}+P_{\mathrm{eff}}$ \label{fig:effective-energy-conditions-wec}]{
    \includegraphics[width=0.48\linewidth]{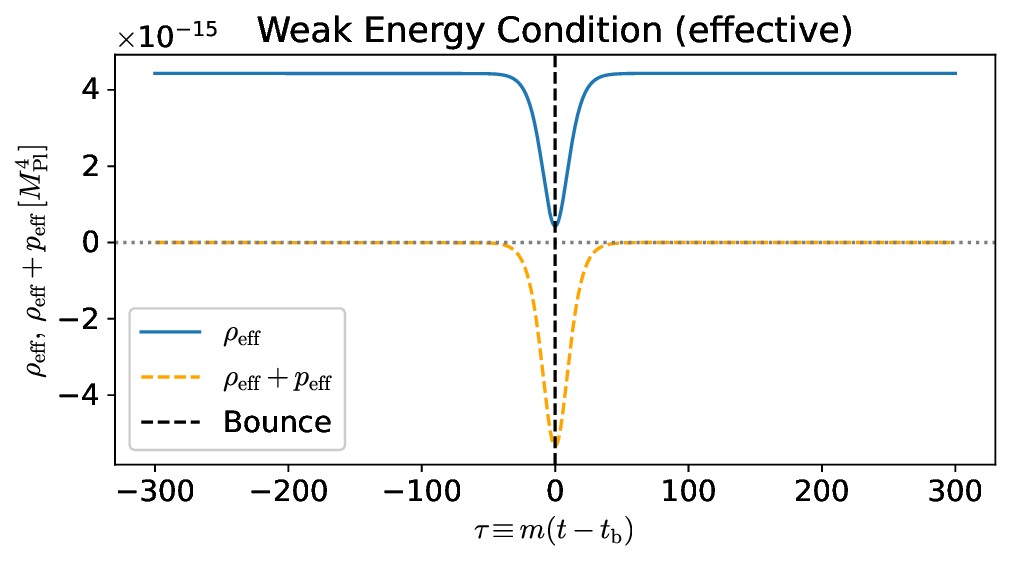}
  }\hfill
  \subfloat[DEC: $\rho_{\mathrm{eff}} -\abs{P_{\mathrm{eff}}}$ \label{fig:effective-energy-conditions-dec}]{
    \includegraphics[width=0.48\linewidth]{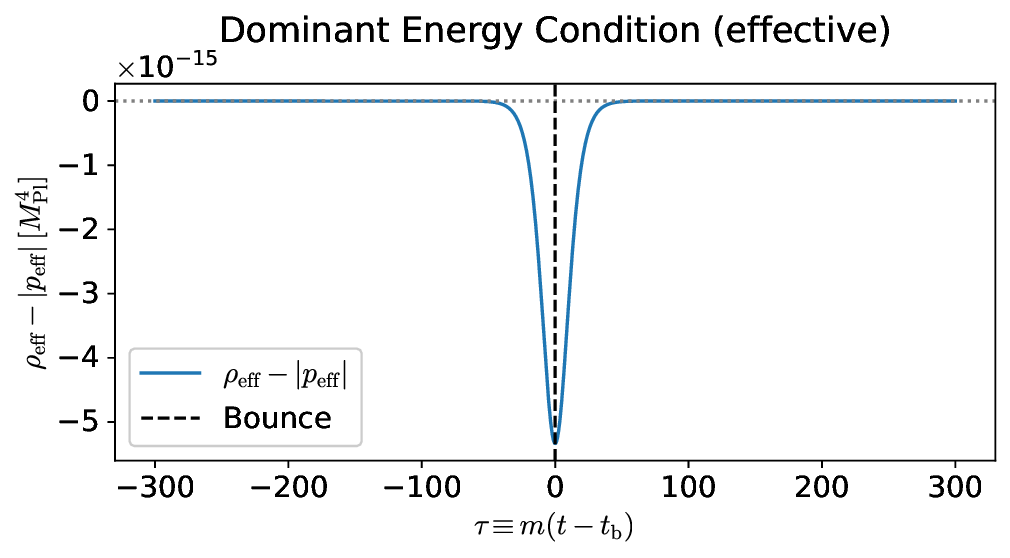}
  }
  \caption{\label{fig:effective-energy-conditions}Evolution of the effective-fluid energy-condition indicators, using the same parameters and initial conditions as in \zcref{fig:scalar_fields}.
  The horizontal axis in all panels is $\tau\equiv m(t-t_{\mathrm{b}})$, and the vertical dashed line marks the bounce ($\tau=0$).
  Panels (a)--(d) correspond to NEC, SEC, WEC, and DEC diagnostics, respectively.
  All quantities are normalized as indicated on the plot axes.}
\end{figure}

The effective-fluid indicators are collected in \zcref{fig:effective-energy-conditions}.
The effective energy density $\rho_{\mathrm{eff}}$ remains nonnegative throughout the displayed range, while the combinations that control NEC, SEC, and DEC exhibit characteristic sign changes near the bounce.
In particular, $\rho_{\mathrm{eff}}+P_{\mathrm{eff}}$ becomes negative around $\tau=0$, corresponding to $\dot H > K/a^2$ via \zcref{eq:nec-eff} and implying NEC violation for the effective cosmic fluid during the bounce phase.
The quantity $\rho_{\mathrm{eff}}+3P_{\mathrm{eff}}$ is also negative, signaling SEC violation; in the present solution, it remains negative across the plotted interval.
Furthermore, the DEC becomes negative in a neighborhood of the bounce, indicating that the effective description also violates DEC there.
Consequently, when the gravitational field is rewritten in terms of an effective source in general relativity, all four standard energy conditions are violated in the bounce regime.

\begin{table}[tbp]
\caption{\label{tab:energy_conditions_bounce}Summary of energy conditions for the effective cosmic fluid and the scalar field at the bounce ($\tau=0$).}
\begin{ruledtabular}
\begin{tabular}{lcc}
Energy condition
  & Scalar field $T^{(\varphi)}_{\mu\nu}$
  & Effective cosmic fluid $T^{(\mathrm{eff})}_{\mu\nu}$\\
\hline
NEC: $\rho + P \ge 0$
  & \OK
  & \NO \\
SEC: $\rho + 3P \ge 0$
  & \NO
  & \NO \\
WEC: $\rho \ge 0,\ \rho + P \ge 0$
  & \OK
  & \NO \\
DEC: $\rho \ge 0,\ \rho \ge \abs{P}$
  & \OK
  & \NO \\
\end{tabular}
\end{ruledtabular}
\end{table}

We emphasize that the above statements concern background (homogeneous) energy-condition diagnostics.
A dedicated perturbation analysis is required to assess possible ghost and gradient instabilities in this model.

\section{\label{sec:conclusions}Conclusions}

In this paper, we have explored the energy conditions realized by bouncing solutions in quadratic curvature gravity with a non-minimally coupled scalar field in detail.
We have obtained the background dynamics by numerically solving the Friedmann equations \eqref{eq:Friedmann-1-mod} and \eqref{eq:Friedmann-2-mod} together with the equations of motion for the scalar field \eqref{eq:EOM-psi-FLRW} and \eqref{eq:EOM-phi-FLRW}.
The resulting solution has exhibited a smooth transition from contraction ($H<0$) to expansion ($H>0$) at $\tau=0$, with $\dot H>0$ at the bounce and a non-zero minimum of the scale factor.

To analyse the energy conditions, we have employed two complementary descriptions.
First, we have evaluated the energy conditions for the scalar field alone by using \zcref{eq:rhophi-Pphi}.
In this case, NEC, WEC, and DEC remain satisfied, whereas SEC is violated around the bounce, as summarized in \zcref{tab:energy_conditions_bounce}.
Second, by rewriting the background equations in the Einstein-like form \zcref{eq:Einstein-eff}, an effective cosmic fluid characterized by $(\rho_{\mathrm{eff}},P_{\mathrm{eff}})$ has been defined.
The numerical indicators have demonstrated that, during the bounce phase, this effective cosmic fluid has violated NEC, SEC, WEC, and DEC.
This contrast has clarified that the scalar field has satisfied the NEC, WEC, and DEC throughout the evolution while violating the SEC during the bounce, whereas the Einstein-like effective cosmic fluid has violated the NEC, SEC, WEC, and DEC near the bounce, showing that the required “exotic” behavior has been encoded in the gravitational sector.

Energy conditions were originally formulated in general relativity as sufficient conditions or the focusing of timelike and null geodesic congruences through the Raychaudhuri equation.
In higher-curvature and more general $f(R)$-type theories, however, their implementation is more nuanced: the conditions may be applied directly to the matter sector, or the modified field equations may be translated into an effective energy-momentum tensor, and the resulting interpretation can depend on the chosen formulation and frame~\cite{Santos:2007bs,Bertolami:2009cd,Capozziello:2018wul}.
For a comprehensive review of $f(R)$ theories of gravity and their viability criteria, see Ref.~\cite{Sotiriou2008}.
From this viewpoint, the fact that the effective cosmic fluid violates all four standard energy conditions near the bounce should not be regarded, by itself, as a definitive no-go for the underlying model; rather, it indicates that the higher-curvature sector behaves as an effective NEC-violating component once the dynamics are cast into an Einstein-like form.

It is also worth noting that closely related bounce solutions in $R^2$ gravity have been studied at the level of linear cosmological perturbations.
In particular, in the $R^2$ bounce-to-inflation setup with a non-minimally coupled scalar field, cosmological perturbations were decomposed into scalar, vector, and tensor sectors and analyzed at first order, providing evidence that the dynamics remain stable across the bounce for a large region of parameter space, with the scalaron contributing to the stability~\cite{Daniel:2022ppp}.

A more definitive assessment of viability in the present setup has required a dedicated perturbation analysis tailored to the quadratic-curvature action.
A natural next step is to rewrite the action in the Arnowitt-Deser-Misner (ADM) formalism~\cite{PhysRev.116.1322,Arnowitt:1962hi} and expand it to the second order in scalar and tensor perturbations, extracting the coefficients of the kinetic and gradient terms for the curvature perturbation and gravitational waves.
The sign of the kinetic coefficients diagnoses ghost instabilities, the sign of the effective mass terms controls tachyonic instabilities, and the sign of the propagation speeds squared (e.g.,\ $c_\mathrm{s}^2$ for scalar modes) determines the presence of gradient instabilities~\cite{Gao:2009ht,DeFelice:2010gb,Kobayashi:2016xpl}.
Such an analysis would complement the background-level energy condition diagnostics reported here, providing a fully gauge-invariant analysis of the stability properties of the bounce.

Furthermore, it is important to examine at the level of linear perturbations how violations of energy conditions, in particular the null energy condition identified at the background level, relate to the stability of perturbations. A systematic perturbation analysis also enables the calculation of inflationary observables such as the spectral index of scalar perturbations $n_s$ and the tensor-to-scalar ratio $r$, allowing a direct comparison with recent observational constraints, including those from the sixth data release of ACT~\cite{AtacamaCosmologyTelescope:2025blo}. This provides a clear test of whether the $R^2$-type predictions remain observationally viable once the coupling considered in this work is taken into account. We hope that the results presented here provide a useful starting point for a comprehensive perturbative and observational assessment of the coupled $R^2$ bounce scenario.

\begin{acknowledgments}
The work of Yuki Hashimoto was supported by JST SPRING, Japan Grant Number JPMJSP2190.
Kazuharu Bamba and Sanjay Mandal acknowledge the support by the JSPS KAKENHI Grant Numbers 24KF0100.
The work of Kazuharu Bamba was supported in part by the JSPS KAKENHI Grant Number 25KF0176 and Competitive Research Funds for Fukushima University Faculty (25RK011).
\end{acknowledgments}

\bibliographystyle{apsrev4-2}
\bibliography{refs}

\end{document}